\documentclass[11pt]{article}

\usepackage[preprint]{acl}

\usepackage{times}
\usepackage{latexsym}
\usepackage{amsmath}
\usepackage[T1]{fontenc}

\usepackage[utf8]{inputenc}

\usepackage{microtype}

\usepackage{inconsolata}
\usepackage{makecell}
\usepackage{makecell}
\usepackage{algorithmicx}

\usepackage{graphicx}
\usepackage{tabularx}
\usepackage{booktabs, multirow, xcolor}

\newcommand{\tool}{TraceElephant}

\title{Seeing the Whole Elephant: A Benchmark for Failure Attribution in LLM-based Multi-Agent Systems}

\author{
Mengzhuo Chen$^{\dagger,\ddagger,\S}$,
Junjie Wang$^{\dagger,\ddagger,\thanks{Corresponding authors}}$,
Fangwen Mu$^{\dagger,\ddagger,\S}$,
Yawen Wang$^{\dagger,\ddagger}$,\\
\bfseries
Zhe Liu$^{\dagger,\ddagger}$,
Huanxiang Feng$^{\dagger,\ddagger,\S}$,
Qing Wang$^{\dagger,\ddagger,\footnotemark[1]}$ \\
\\
$^{\dagger}$State Key Laboratory of Complex System Modeling and Simulation Technology \\
$^{\ddagger}$Institute of Software, Chinese Academy of Sciences \\
$^{\S}$University of Chinese Academy of Sciences, Beijing, China
}

\usepackage[linesnumbered,ruled,vlined]{algorithm2e}

\usepackage{float}
\usepackage{threeparttable}
\usepackage{booktabs}
\usepackage{multirow}
\usepackage{epstopdf}
\usepackage{hyperref}
\usepackage{listings}
\usepackage{fancybox}
\usepackage{graphicx}
\usepackage{subcaption}
\usepackage{paralist}
\usepackage{ragged2e}
\usepackage{enumitem}

\usepackage{marvosym}
\usepackage{multirow}
\usepackage{epstopdf}
\usepackage{hyperref}
\usepackage{listings}
\usepackage{fancybox}
\usepackage{graphicx}
\usepackage{subcaption}
\usepackage{subfloat}
\usepackage{cleveref}
\usepackage{paralist}
\usepackage{ragged2e}
\usepackage{enumitem}
\usepackage{pythonhighlight}
\usepackage{algpseudocode}

\usepackage[normalem]{ulem} 
\newcommand\hlg{\bgroup\markoverwith
  {\textcolor{green!10}{\rule[-.5ex]{2pt}{2.5ex}}}\ULon}
\newcommand\hlb{\bgroup\markoverwith
  {\textcolor{blue!10}{\rule[-.5ex]{2pt}{2.5ex}}}\ULon}
\newcommand\hlr{\bgroup\markoverwith
  {\textcolor{red!10}{\rule[-.5ex]{2pt}{2.5ex}}}\ULon}

\usepackage{listings}

\newif\ifFirstMintedPart


\newcommand{\best}[1]{\textbf{#1}}
\newcommand{\second}[1]{\underline{#1}}

\usepackage[table]{xcolor}
\usepackage{soul}
\usepackage{float}
\usepackage{threeparttable}
\usepackage{booktabs}
\usepackage{multirow}
\usepackage{epstopdf}
\usepackage{hyperref}
\usepackage{listings}
\usepackage{fancybox}
\usepackage{graphicx}
\usepackage{subcaption}
\usepackage{paralist}
\usepackage{ragged2e}
\usepackage{enumitem}
\usepackage{balance}

\usepackage{marvosym}
\usepackage{multirow}
\usepackage{epstopdf}
\usepackage{hyperref}
\usepackage{listings}
\usepackage{fancybox}
\usepackage{graphicx}
\usepackage{subcaption}
\usepackage{subfloat}
\usepackage{cleveref}
\usepackage{paralist}
\usepackage{ragged2e}
\usepackage{enumitem}
\usepackage{pythonhighlight}
\usepackage{algpseudocode}

\usepackage{minted}
\usepackage{caption}
\usepackage{xparse}

\newminted{c}{
    linenos, 
    frame=lines, 
    tabsize=4, 
    autogobble,
}

\newminted{java}{
    linenos, 
    frame=lines, 
    tabsize=4, 
    autogobble,
}

\usepackage{listings}

\usepackage[framemethod=TikZ]{mdframed}
\usepackage{tcolorbox}
\makeatletter
\newcommand{\mybox}[1]{%
  \setbox0=\hbox{#1}%
  \setlength{\@tempdima}{\dimexpr\wd0+13pt}%
  \begin{tcolorbox}[boxrule=0.5pt, colback=white, arc=4pt,
      left=6pt,right=6pt,top=6pt,bottom=6pt,boxsep=0pt]
    #1
  \end{tcolorbox}
}
\usepackage{tikz}
\usepackage{pgfplots}
\usetikzlibrary{pgfplots.statistics,calc}
\usepackage{array}
\usepackage{centernot}
\usepackage{xspace}
\usepackage{url}
\usepackage{verbatim}
\usepackage{wrapfig}
\usepackage{tabularx}
\usepackage{bbding}
\usepackage{pifont}
\usepackage{wasysym}
\usepackage{amssymb}

\begin{document}
\maketitle
\begin{abstract}
Failure attribution, i.e., identifying the responsible agent and decisive step of a failure, is particularly challenging in LLM-based multi-agent systems (MAS) due to their natural-language reasoning, nondeterministic outputs, and intricate interaction dynamics. A reliable benchmark is therefore essential to guide and evaluate attribution techniques. Yet existing benchmarks rely on partially observable traces that capture only agent outputs, omitting the inputs and context that developers actually use when debugging.
We argue that failure attribution should be studied under full execution observability, aligning with real-world developer-facing scenarios where complete traces, rather than only outputs, are accessible for diagnosis.
To this end, we introduce {\tool}, a benchmark designed for failure attribution with full execution traces and reproducible environments.
We then systematically evaluate failure attribution techniques across various configurations. 
Specifically, full traces improve attribution accuracy by up to 76\% over a partial-observation counterpart, confirming that missing inputs obscure many failure causes.
{\tool} provides a foundation for follow-up failure attribution research, promoting evaluation practices that reflect real-world debugging and supporting the development of more transparent MASs.

\end{abstract}

\section{Introduction}
\label{sec_introduction}
If there is one constant in the evolution of software, it is the persistent occurrence of failures \cite{charette2005software}. 
When it does, the critical first step is to assign responsibility for the failure to a specific component, i.e., failure attribution or localization, which enables developers to focus debugging efforts, guide the design of patches or architectural improvements.
Traditional techniques for this task, ranging from statistical debugging \cite{zheng2006statistical} and delta debugging \cite{misherghi2006hdd} to more recent learning-based approaches \cite{wong2016survey,zou2019empirical}, operate under the assumption that system states are discrete, executions are traceable, and component behaviors are largely deterministic.

The rise of LLM-based multi-agent systems (MASs) fundamentally challenges these assumptions and complicates the attribution problem.
In these systems, complex tasks are decomposed and coordinated across multiple agents whose primary reasoning and communication medium is natural language \cite{guo2024large,li2024survey}. 
This introduces two layers of complexity for fault attribution. 
First, the problem-solving process in MASs often involves complex interactions among multiple LLM-powered agents, between agents and external tools, and within the internal reasoning processes of the LLMs themselves. These interactions complicate system logs, challenging the interpretation of system behavior and hindering rapid root cause identification. 
Second, the system actions and their resulting states are recorded in natural language within the log. 
The inherent ambiguity of natural language further impedes the precise characterization of operations and states.

To address the challenge of fault attribution in MAS, several technical approaches have been proposed, such as ECHO \cite{banerjee2025ECHO}, AgenTracer \citep{zhang2025agentracer}, GraphTracer \citep{zhang2025graphtracer}, and FAMAS \cite{ge2025spectrumanalysis}.
On another front, to effectively evaluate how these techniques perform in real-world MAS settings, the foundation lies in having benchmarks that accurately reflect authentic fault attribution scenarios. 

\begin{figure*}[t]
    \centering
    \includegraphics[width=0.98\textwidth]{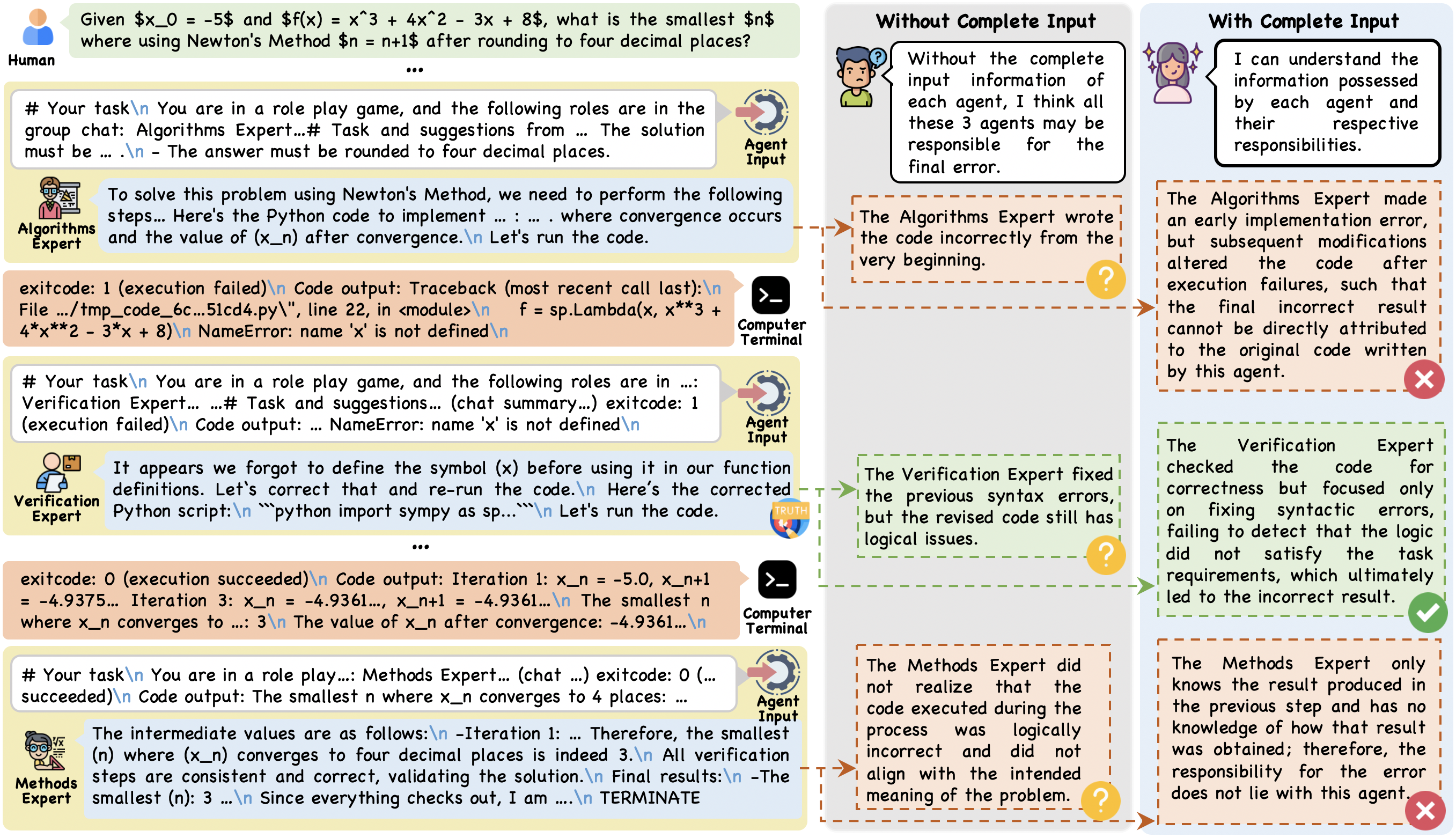}
    \caption{A failure case (from Who\&When benchmark) illustrating the limitation of partial observability. 
    When only agent outputs are visible and critical inputs are absent, localizing the decisive failure step becomes difficult.}
    \label{fig:who_when_incomplete}
\end{figure*}

To the best of our knowledge, Who\&When \cite{zhang2025agent} is currently the only benchmark specifically designed for failure attribution in LLM-based MAS. It provides partially observable execution traces that include only the agents' outputs, without the corresponding inputs such as original task instructions, prompts, or contextual messages. This restricted setting can be suitable for evaluating attribution methods under certain black-box scenarios, where internal inputs or intermediate states are inaccessible (e.g., debugging deployed agents on external platforms).

However, in many practical developer-facing debugging scenarios, developers typically have access to substantially richer execution information, including task instructions, prompts, intermediate messages, tool invocations, and environment states. More importantly, we view failure attribution as a developer-centric task conducted in the context of system debugging and iterative refinement, with the explicit goal of producing actionable insights for repairing and improving MASs. Achieving this goal critically depends on access to the inputs, intermediate states, and environment interactions that drive agent behavior. In contrast, black-box settings, where only outputs are observable, leave many failures ambiguous and thus offer limited practical value for guiding system fixes. For instance, our analysis of the 184 failure cases in Who\&When benchmark shows that in at least 21\% of instances, developers cannot reliably perform failure attribution using output-only logs. 
We illustrate this in Figure~\ref{fig:who_when_incomplete} with an adapted case from Who\&When, where we restored the missing inputs by re-running the original system. Under partial observability, the failure cause remains ambiguous, whereas with full inputs, the ambiguity is largely resolved, enabling precise step-level localization.

To fill this gap, we construct {\tool}, the first benchmark specifically designed for developer-facing failure attribution in LLM-based MASs by ``seeing the whole elephant'', i.e., providing access to the complete execution narrative of an MAS.
Although we use MAS as the main terminology, the attribution unit in {\tool} is a functional component rather than necessarily a separate agent entity, so the benchmark also applies to single-agent scaffold systems whose planning, orchestration, and tool-use modules make distinct decisions.
Compared with existing benchmarks in this field (i.e., Who\&When), {\tool} differs in two key aspects: (1) {\tool} collects step-by-step execution traces from multiple representative agentic systems, where each trace records agent-level actions, natural language inputs and outputs, tool and environment interactions, agent configurations and system architecture, and (2) it accompanies each trace with a reproducible execution environment, enabling controlled re-execution, state inspection, and interactive hypothetical debugging queries (e.g., ``what if this agent had received different input?'').

Building upon this benchmark, we evaluate the automated failure attribution techniques under various configurations. 
The experimental results demonstrate that with complete trace information, it achieves an average attribution accuracy of 65.9\% at the agent level and 30.3\% at the step level. 
This represents an improvement of 22\% in agent-level accuracy and 76\% in step-level accuracy over the performance on output-only traces (i.e., similar to Who\&When benchmark), underscoring the critical role of full observability.
With the running environment, the step-level accuracy can further improve by 10\%.
Our analysis further reveals that step-level attribution is more sensitive to missing information than agent-level attribution, highlighting the finer-grained, context-dependent nature of localizing the precise failure step.
Moreover, performance can also vary across different MAS architectures, agent types, and step positions, indicating the need for architecture-aware attribution approaches.
We also provide actionable implications and takeaways for developing more effective attribution techniques and designing more debuggable MASs.

On one hand, this work  highlights the necessity of full trace information for reliable fault attribution, indicating that developers should, whenever possible, incorporate all accessible execution details when performing this task. 
On the other hand, it also suggests that the field would benefit from more diverse benchmarks to evaluate attribution techniques from multiple perspectives, thereby 
contributing to a more solid and cumulative understanding of failure attribution in MAS.

The contributions of this work are as follows.
\begin{itemize}
\item We are the first to study failure attribution of LLM-based MASs from a developer-facing scene, where full execution traces and reproducible execution environment are available. 
\item We develop {\tool}\footnote{\url{https://github.com/TraceElephant/TraceElephant}}, a failure attribution benchmark that instantiates the above paradigm by providing a collection of annotated execution traces. It consists of 220 failure traces collected from three representative agentic systems, including both multi-agent orchestration and a single-agent tool-centric scaffold, with each annotated with the responsible component and decisive failure step.
\item We conduct extensive experiments on {\tool}, examining how failure attribution behaves under various configurations, and providing related implications. 
\end{itemize}

\section{Problem Definition}
\label{sec:problem_definition}
We consider an LLM-based multi-agent system (MAS) composed of a finite set of functional components $\mathcal{A} = \{a_1, \dots, a_N\}$ collaboratively performing a task $\tau$.
These components may correspond to explicit agents in a classical MAS, or to functional modules in a single-agent scaffold, such as planning, orchestration, and tool-use components.
The system executes in discrete steps under a turn-based protocol, where exactly one component is selected to act at each step. At step $t$, the acting component $a(t)$ receives input $x_t$ and produces output $y_t$. The execution trace at step $t$ is recorded as
\[
o_t = \big(x_t,\, y_t,\, a(t),\, \textit{step\_id}_t,\, \textit{agent\_id}_t \big).
\]

The task outcome is determined by the complete execution trace. In case of failure, we aim to attribute responsibility at both the step and agent level: step-level attribution identifies the earliest point at which failure becomes inevitable, while agent-level attribution identifies the component responsible for the failure at that step.
This definition follows a role-aware and recoverability-aware principle rather than a purely chronological notion of the first visible mistake.
If an upstream mistake remains recoverable because a later verifier or orchestrator is explicitly responsible for checking and correcting it, the decisive failure step is attributed to the point where that recovery opportunity is missed.
For example, if an agent produces a hallucinated fact at step $t$ but a verifier at step $t+2$ is expected to detect such errors and fails to do so, the failure is attributed to the verifier step, because the system remains recoverable until that responsibility is not fulfilled.
For the task of failure attribution in MAS, the goal is to determine both the step-level and agent-level responsibility given the full failure trace.

For a full formalization, including definitions, notation, and equations, refer to Appendix~\ref{app:problem_definition}.

\begin{figure*}[!th]
\centering
\vspace{-0.1in}
\includegraphics[width=\textwidth]{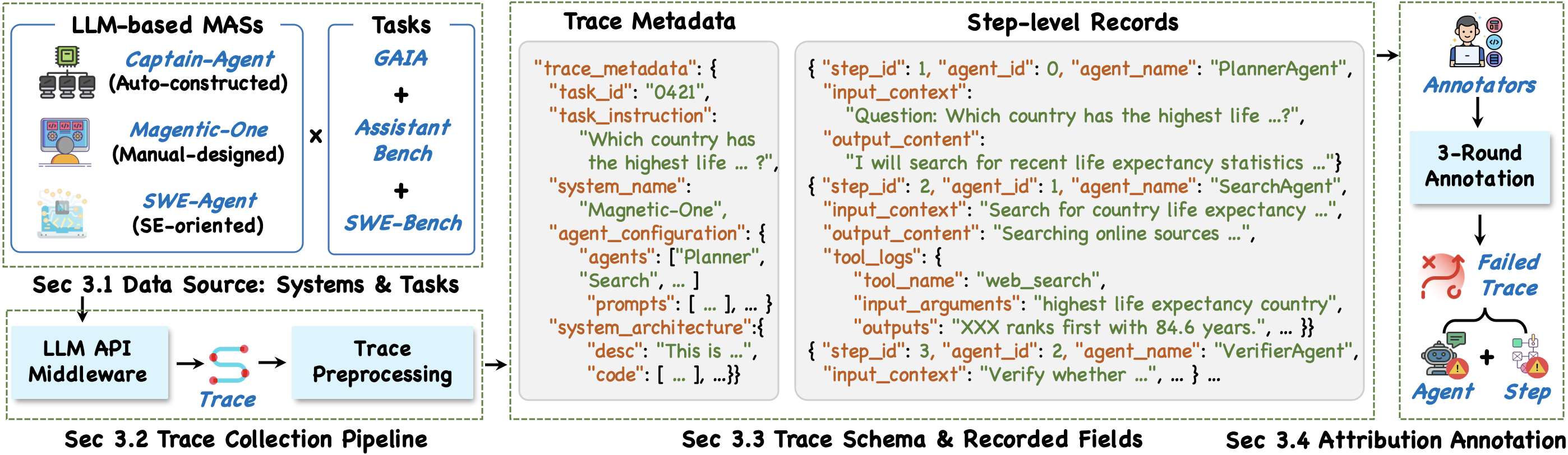}
\caption{Overview of {\tool}.}
\label{fig:overview}
\vspace{-0.1in}
\end{figure*}

\section{Benchmark Construction}
\label{sec:dataset}
We construct {\tool}, a benchmark for failure attribution in LLM-based MAS.
Each instance is grounded in an executable MAS and is associated with a fully observable execution trace, as well as the annotated failure-responsible agent and decisive failure step.

\subsection{Data Sources: Systems and Tasks}
\label{sec:dataset_sources}
{\tool} collects execution traces from three representative agentic systems across various tasks, aligning task design with each system's intended capabilities, as demonstrated in Table \ref{tab:trace_stats}.
These systems cover dynamically assembled agent teams, fixed-role multi-agent orchestration, and a single-agent tool-centric scaffold, allowing the same attribution formulation to apply across both explicit multi-agent interactions and scaffolded single-agent workflows.
Detailed descriptions of the systems, task sources, run configurations are provided in Appendix \ref{app:dataset_sources}.

\begin{table}[htb]
\centering
\small
\begin{tabularx}{\columnwidth}{l X c c}
\toprule
\textbf{System} & \textbf{Task Source} & \textbf{\# Traces} & \textbf{\# Failed} \\
\midrule
Captain-Agent & GAIA               & 126 & 73 \\
Captain-Agent & AssistantBench     & 21 & 12 \\
Magentic-One  & GAIA               & 119 & 74 \\
Magentic-One  & AssistantBench     & 30 & 17 \\
SWE-Agent     & SWE-Bench      & 84 & 44 \\
\midrule
\textbf{Total} & All & 380 & 220 \\
\bottomrule
\end{tabularx}
\caption{Overview of execution traces in {\tool}.}
\label{tab:trace_stats}
\vspace{-0.15in}
\end{table}

\subsection{Trace Collection Pipeline}
\label{sec:trace_adapter}

{\tool} adopts an automated collection pipeline designed to capture complete execution traces while preserving the original structure of agent interactions. At its core, a lightweight LLM API middleware  transparently intercepts and captures all LLM requests, responses, and subsequent tool interactions without modifying the original agent implementations.
The raw traces then undergo targeted pre-processing, extracting only basic attributes such as agent names and step types, to maintain maximal fidelity to the original execution flow. 
More details are provided in Appendix \ref{app:trace_adapter}.

\subsection{Trace Schema and Recorded Fields}
\label{sec:trace_fields}
Each instance in {\tool} benchmark consists of a complete execution trace generated by running a multi-agent system on a given task, as well as the executable system associated with the runnable code and configurations for this specific trace instance.

\textbf{(1) Trace metadata.}
Each execution trace is associated with a set of trace-level metadata that describes the execution context as a whole.
These metadata fields include:
(1) \textit{task\_id} and \textit{task\_instruction}, indicating the task being solved;
(2) \textit{system\_name}, identifying the multi-agent system that produced the trace; 
(3) \textit{agent\_configuration}, the runtime setup defining the agent roster, prompts, and toolset; and 
(4) \textit{system\_architecture}, the design documentation and implementation code defining the system's structural blueprint.

\textbf{(2) Step-level records.}
An execution trace is composed of an ordered sequence of steps, where each step corresponds to a single agent action.
For each step, {\tool} records a set of observable fields, organized into input and output fields, that are directly exposed by the running system.

\textbf{\textit{(i) Step input fields.}}
They capture all information provided to an agent when an action is executed.
Specifically, the recorded input fields include:
(1) \textit{step\_id}, a unique identifier indicating the global execution order of the step;
(2) \textit{agent\_id} and \textit{agent\_name}, identifying the agent responsible for the action;
(3) \textit{input\_context}, which stores the complete input information supplied to the agent.
This field typically includes the original task instruction, intermediate messages exchanged among agents, and any system-constructed contextual information available at execution time.

\textbf{\textit{(ii) Step output fields.}}
They record the observable results produced by the agent at the current step.
These fields include:
(1) \textit{output\_content}, which stores the response generated by the agent.
(2) \textit{tool\_logs}, which (when available) records raw interaction logs with external tools or environments when such interactions occur.
Each tool log typically includes the tool name, input arguments, outputs, and execution status.

All execution traces are stored in a structured JSON format. Each trace is represented as a JSON object containing trace-level metadata and an ordered list of step records, within which input and output fields are stored as nested objects, preserving all content in its raw, unprocessed form.
We provide an example trace in Appendix \ref{app:trace_example}.

\subsection{Failure Attribution Annotation}
The annotation aims at acquiring failure attribution labels, i.e., (1) the component primarily responsible for the failure, and (2) the execution step where the failure originates.
These labels are obtained through a multi-round expert annotation process designed to ensure reliability.
On the first-round independent annotations, Krippendorff's alpha is 0.72 for agent-level labels and 0.64 for step-level labels, indicating substantial agreement for agent attribution and moderate-to-substantial agreement for step identification.
Afterward, uncertain cases are jointly reviewed and finalized through expert consensus.
Detailed annotation protocols are provided in Appendix ~\ref{app:annotation}.

\section{Failure Attribution Evaluation}
\label{sec:evaluation}

\begin{table*}[t]
\centering
\small
\setlength{\tabcolsep}{4pt}
\renewcommand{\arraystretch}{1}
\vspace{-0.1in}
\begin{tabular}{llcccccccccc}
\toprule
\multirow{3}{*}{\raisebox{-1.2\height}{\textbf{System}}} &
\multirow{3}{*}{\raisebox{-1.2\height}{\textbf{Ground Truth}}} &
\multicolumn{8}{c}{\textbf{Static Configurations}} &
\multicolumn{2}{c}{\textbf{Dynamic Config.}} \\
\cmidrule(lr){3-10} \cmidrule(lr){11-12}

& &
\multicolumn{2}{c}{\textbf{All-at-Once}} &
\multicolumn{2}{c}{\textbf{Binary Search}} &
\multicolumn{2}{c}{\textbf{Step-by-Step}} &
\multicolumn{2}{c}{\textbf{Static Agentic}} &
\multicolumn{2}{c}{\textbf{Dynamic Agentic}} \\
\cmidrule(lr){3-4}
\cmidrule(lr){5-6}
\cmidrule(lr){7-8}
\cmidrule(lr){9-10}
\cmidrule(lr){11-12}

& &
\textbf{Agent} & \textbf{Step} &
\textbf{Agent} & \textbf{Step} &
\textbf{Agent} & \textbf{Step} &
\textbf{Agent} & \textbf{Step} &
\textbf{Agent} & \textbf{Step} \\
\midrule

\multirow{2}{*}{\textbf{Captain-Agent}}
& \textbf{w/} Ground Truth  & 64.7 & 29.4 & 25.9 & 14.1 & 57.7 & 22.4 & \second{67.1} & \second{30.6} & \best{68.2} & \best{32.9} \\
& \textbf{w/o} Ground Truth & \best{63.5} & 22.4 & 24.7 & 9.4 & 44.7 & 17.7 & 58.8 & \second{24.7} & \second{61.2} & \best{25.9} \\

\midrule

\multirow{2}{*}{\textbf{Magentic-One}}
& \textbf{w/} Ground Truth  & 58.2 & 25.2 & 38.5 & 13.2 & 63.7 & 20.9 & \second{67.0} & \second{30.8} & \best{68.1} & \best{33.0} \\
& \textbf{w/o} Ground Truth & 56.0 & 23.1 & 37.4 & 8.8 & 60.4 & 15.4 & \second{61.5} & \second{26.4} & \best{61.5} & \best{27.5} \\

\midrule

\multirow{2}{*}{\textbf{SWE-Agent}}
& \textbf{w/} Ground Truth  & \best{63.6} & \second{29.6} & 52.3 & 11.4 & \second{61.4} & 6.8 & \best{63.6} & \second{29.6} & \best{63.6} & \best{34.1} \\
& \textbf{w/o} Ground Truth & 54.6 & 22.7 & 50.0 & 9.1 & \second{56.8} & 4.6 & \second{56.8} & \second{27.3} & \best{59.1} & \best{29.6} \\
\midrule

\multirow{2}{*}{\textbf{All-avg}}
& \textbf{w/} Ground Truth & 62.2 & 28.1 & 38.9 & 12.9 & 60.9 & 16.7 & \second{65.9} & \second{30.3} & \best{66.7} & \best{33.3} \\
& \textbf{w/o} Ground Truth & 58.0 & 22.7 & 37.4 & 9.1 & 54.0 & 12.5 & \second{59.1} & \second{26.1} & \best{60.6} & \best{27.6} \\

\bottomrule
\end{tabular}

\caption{Performance (i.e., Agent-level accuracy and Step-level accuracy) comparison of failure attribution techniques across different agent systems under \textit{with or without ground truth} scenario. We use bold to indicate the highest value, while underline indicating the second highest value. }
\label{tab:main_results}
\end{table*}

\begin{table}[t]
\centering
\small
\setlength{\tabcolsep}{6pt}
\renewcommand{\arraystretch}{1.1}
\begin{tabular}{lcc}
\toprule
\textbf{Observability Configurations} &
\textbf{Agent} &
\textbf{Step} \\
\midrule

\textbf{All-at-Once}                  & \textbf{0.62} & \textbf{0.28} \\
\quad w/o metadata                    & 0.55 & 0.21 \\
\quad w/o input                       & 0.54 & 0.18 \\
\quad w/o metadata \& input           & 0.51 & 0.16 \\

\midrule

\textbf{Static Agentic}               & \textbf{0.66} & \textbf{0.30} \\
\quad w/o metadata                    & 0.57 & 0.23 \\
\quad w/o input                       & 0.56 & 0.19 \\
\quad w/o metadata \& input           & 0.54 & 0.17 \\

\bottomrule
\end{tabular}
\caption{Ablation study on observability configurations.}
\vspace{-0.1in}
\label{tab:observability-ablation}
\end{table}

\subsection{Evaluation Design}
\label{sec:eval_settings}

\textbf{Observability Configurations.} We evaluate failure attribution performance under two complementary observability configurations reflecting practical debugging workflows. (1) \textbf{\textit{Static}}: Attribution is performed using the complete execution trace (including metadata, inputs, and outputs fields as shown in Section \ref{sec:trace_fields}). (2) \textbf{\textit{Dynamic}}: In addition to the static trace, a replayable execution environment is provided, enabling controlled re-execution and counterfactual probing to verify or refine candidate attributions.
We additionally utilize several variants of static configuration 
to further evaluate the performance under different levels of observability.

\textbf{Attribution techniques.} We utilize five commonly-used techniques. 
There are three LLM-based prompting techniques, i.e., \textbf{\textit{All-at-Once, Binary Search, Step-by-Step}}, which differ in how the trace is provided to the LLM.
There are two agent-based techniques. 
\textbf{\textit{Static Agentic}} adopts mini-SWE-agent\footnote{2.4k stars as of January 2026. https://github.com/SWE-agent/mini-swe-agent/.}, which can navigate the trace information, retrieve related fields as needed, and make a conclusion gradually.
\textbf{\textit{Dynamic Agentic}} first proposes candidate failure attributions derived based on static agentic technique, with the candidate steps and agents. The method then re-runs the system from the corresponding execution point and issues counterfactual checks.
Default setting of the following evaluations is Static Agentic.

\textbf{Application Scenarios.} Following existing studies \cite{zhang2025agent}, experiments are conducted both \textbf{\textit{With ground truth}} (use this, if not explicitly stated) and \textbf{\textit{Without ground truth}} scenarios, simulating different debugging contexts. 

\textbf{Evaluation Metrics.} Predictions are evaluated for both responsible agent and decisive step, and we use \textbf{\textit{agent-level accuracy}} and \textbf{\textit{step-level accuracy}}, following existing studies \cite{zhang2025agent, zhang2025graphtracer}.

More details about the evaluation design are provided in Appendix \ref{app:evaluation_design}.

\begin{figure}[!tb]
\centering
\includegraphics[width=0.47\textwidth]{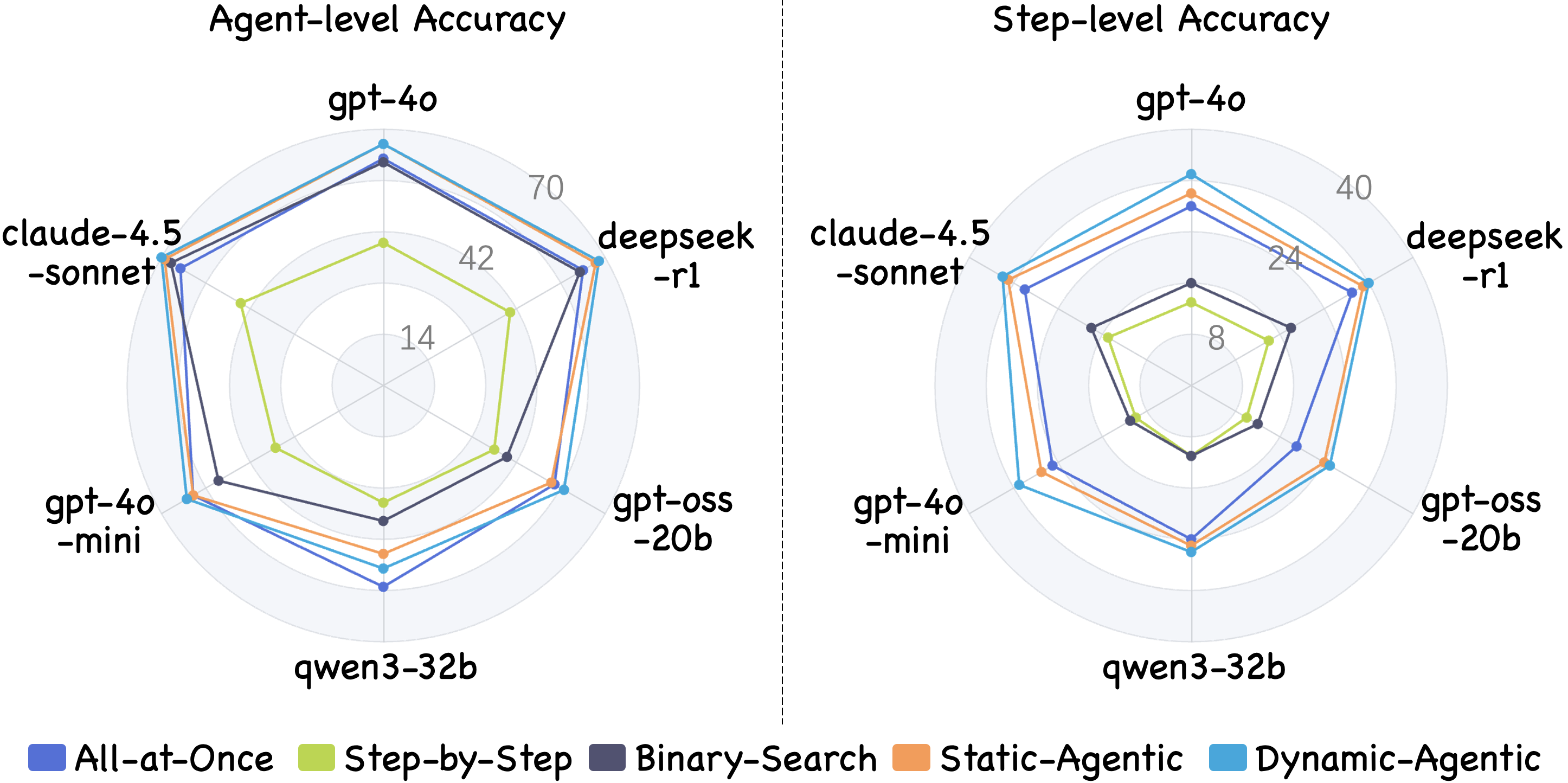}
\caption{Comparison under different backbone LLMs.}
\label{fig:radar}
\vspace{-0.1in}
\end{figure}

\subsection{Results and Analysis}
\label{sec:main_results}

\subsubsection{Performance Across Configurations}

\textbf{(1) Static vs. Dynamic Attribution Performance.}
Generally speaking, dynamic configuration can achieve the highest performance, especially for step-level accuracy, and within static configurations, the agentic technique is the best in most experimental settings.

As demonstrated in Table \ref{tab:main_results}, with ground truth, failure attribution performance can reach an average of 33.3\% and 30.3\% step-level accuracy respectively in dynamic and static configuration. 
For agent-level accuracy, these figures are 66.7\% and 65.9\% respectively.
The dynamic method improves step-level attribution by 10\% due to actively verifying candidate failure steps through controlled re-execution and counterfactual probing. This process helps filter out spurious candidates identified from static traces, thereby refining the attribution.
Agent-level accuracy sees limited gains because it depends more on understanding agent roles and coordination logic, i.e., information already largely available in complete static traces.

Among all static configurations, agentic technique achieves the highest performance in almost all cases, which is likely because it allows to better trace responsibility through the chain of analysis and tool use that characterizes MAS failures.
Furthermore, among other static configuration, All-at-Once performs relatively better than others. 
We conduct further analysis in Section \ref{sec_result_whowhen}.

\textbf{(2) Ablation Study.}
Table \ref{tab:observability-ablation} presents the ablation results for two representative techniques, while others show similar trend and are omitted for brevity. 
Full observability is essential for accurate failure attribution, and the absence of either metadata or input fields leads to noticeable performance degradation.
Besides, step-level attribution is more sensitive to missing information than agent-level attribution, i.e., 76\% vs. 22\% accuracy drop, underscoring the finer-grained and more context-dependent nature of identifying the failure step.

\textbf{(3) Effect of Backbone LLMs.}
Figure \ref{fig:radar} demonstrates the performance in terms of different backbone LLMs. 
Broadly speaking, Claude-4.5-Sonnet, DeepSeek-R1, and GPT-4o exhibit relatively strong performance in both agent-level and step-level accuracy, likely due to their advanced reasoning architectures and strong contextual understanding. 
In contrast, Qwen3-32B and GPT-OSS-20B show weaker performance, which may be attributed to their smaller parameter scales and consequently limited capacity for sustained multi-step reasoning and fine-grained causal tracing. 

\textbf{(4) With vs. Without Ground Truth.}
Performance consistently declines across all systems and techniques when ground truth is unavailable. This confirms that access to a reference outcome (e.g., a correct answer or test pass/fail signal) provides crucial guidance for attribution, especially for finer-grained step attribution.
Agentic methods (especially Dynamic Agentic) show relatively smaller performance degradation without ground truth. 
This suggests their interactive validation (replay and counterfactual checks) can partially compensate for the missing reference signal by testing behavioral hypotheses.

\subsubsection{Failure \& Attribution Patterns}
\label{sec_failure_patterns}
\begin{figure}[!t]
\centering
\includegraphics[width=0.47\textwidth]{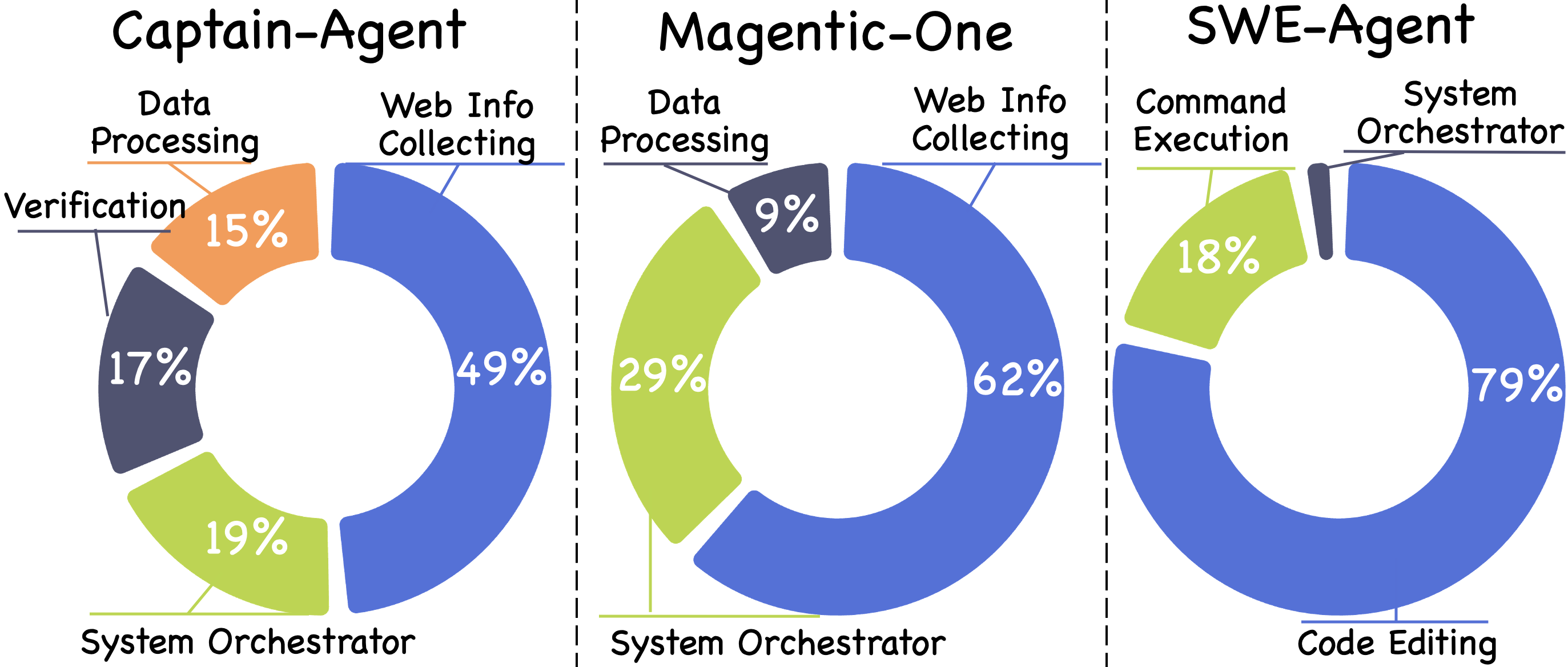}
\vspace{-0.05in}
\caption{Distribution of failure  agent in {\tool}.}
\label{fig:failure_agent}
\vspace{-0.05in}
\end{figure}

\textbf{(1) Failure Responsible Agents.} 
Figure \ref{fig:failure_agent} illustrates the distribution of failure-prone agent types in our benchmark.
Agents responsible for interacting with external environments or performing concrete operations are most prone to errors (almost over 50\% of the failures), i.e., agents handling web information collection and browsing in CaptainAgent and Magentic-One, agent directly editing code in SWE-Agent.
This might be because these actions depend on dynamic, often noisy external systems (APIs, websites, file systems), where malformed requests, parsing errors, or unexpected outputs can easily occur. 
The orchestrator/planner agents also represent a non-negligible (18-29\%) source of failures.
Their mistakes often stem from incorrect task decomposition, sub-optimal agent selection, or flawed coordination logic-errors that may not manifest immediately but can propagate and amplify throughout the execution.

\begin{figure}[!t]
\centering
\includegraphics[width=0.45\textwidth]{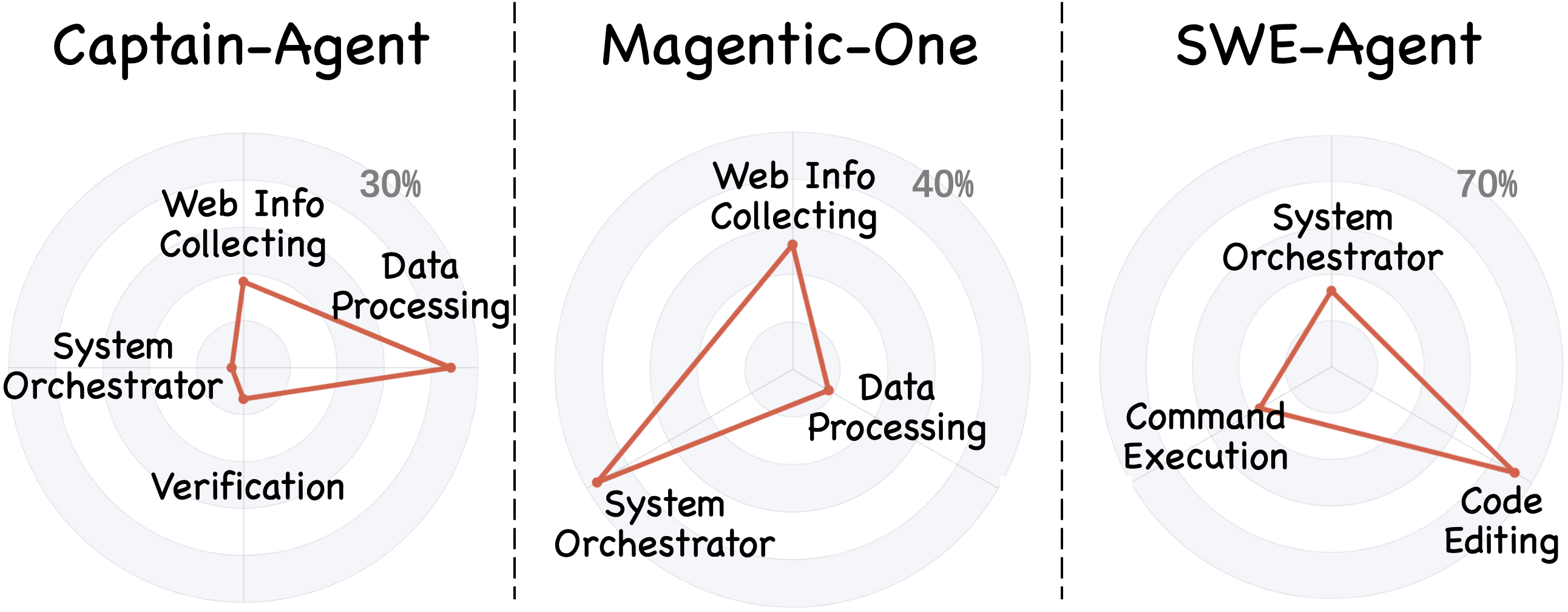}
\vspace{-0.05in}
\caption{Fine-grained agent-level accuracy.}
\label{fig:agent-level}
\vspace{-0.15in}
\end{figure}

\textbf{(2) Agent-level Accuracy.}
Figure~\ref {fig:agent-level} further breaks down the predicted agent-level accuracy for these key agent types.
For Captain-Agent, data processing agents show high prediction accuracy (53\%), likely because their errors (e.g., CSV parsing or script execution failures) leave clear, tool-mediated traces, and web-related agents achieve moderate accuracy (22\%), as search API outputs are often ambiguous in natural language.
For Magentic-One, the Orchestrator attains higher accuracy (38\%) than in Captain-Agent, possibly because its failures frequently arise during visible mid-process interventions, such as re-planning after a browser search stalls, making its responsibility more discernible from output logs.

\begin{figure}[!t]
\centering
\vspace{0.1in}
\includegraphics[width=0.47\textwidth]{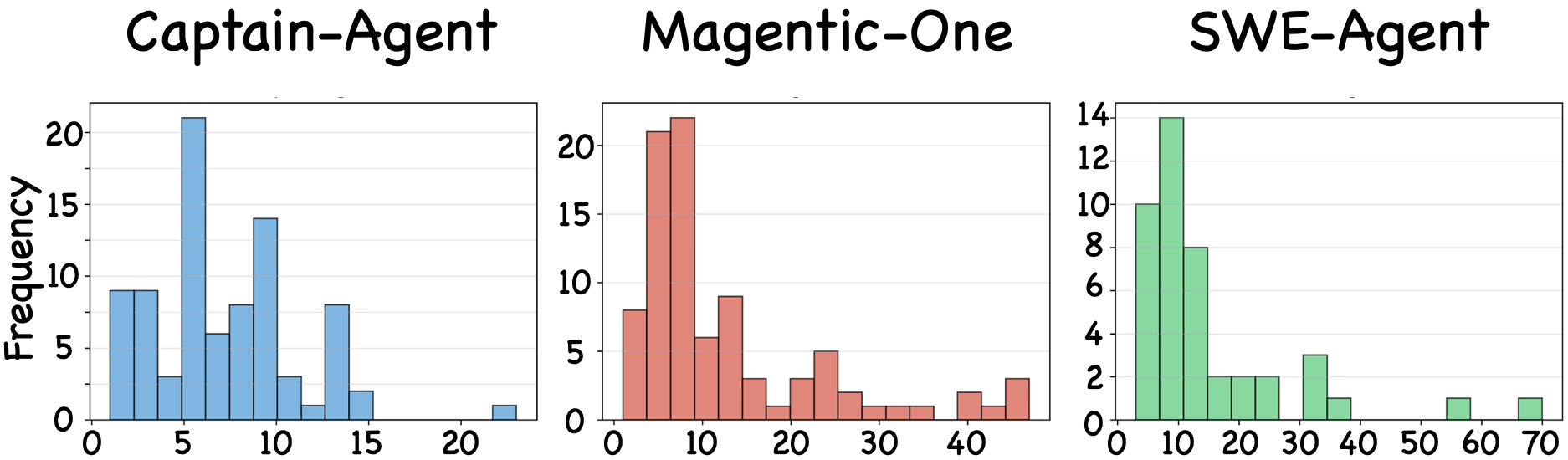}
\caption{Distribution of failure step in {\tool}.}
\label{fig:failure_step}
\end{figure}

\textbf{(3) Decisive Failure Steps.}
Figure \ref{fig:failure_step} shows the distribution of failure steps over the execution timeline in our benchmark.
Failures in the automatically generated system (i.e., Captain-Agent) are dispersed.
Since the system constructs and coordinates agent teams on-the-fly for each task, errors can be introduced at multiple points: during agent selection, iterative planning, inter-agent coordination, or tool-calling.
Failures in manually crafted systems (i.e., Magentic-One and SWE-Agent) are heavily concentrated in the early steps, tightly linked to the initial task planning and routing decisions made by their central orchestrators.

\begin{figure}[!t]
\centering
\includegraphics[width=0.47\textwidth]{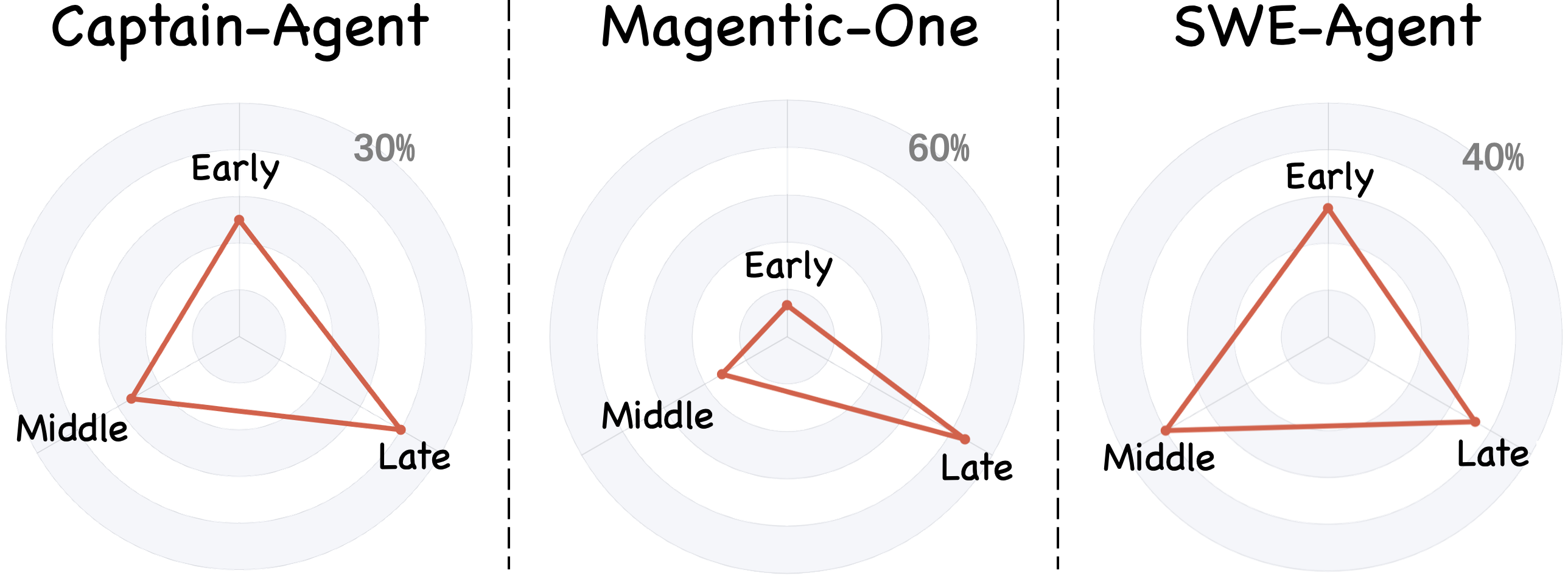}
\caption{Fine-grained step-level accuracy.}
\label{fig:step-level}
\vspace{-0.1in}
\end{figure}

\textbf{(4) Step-level Accuracy.}
Figure~\ref{fig:step-level} demonstrates the step-level accuracy for the ground-truth failure steps divided equally into three segments based on their chronological order.
In Captain-Agent and SWE-Agent, step-level accuracy remains relatively stable across all phases, while for Magentic-One, the accuracy is remarkably low in the early phase (8\%), moderate in the middle (19\%), and high in the late phase (52\%). 
This suggests that failures occurring early in the execution are especially difficult to attribute, whereas those in later stages are more easily identified.
A plausible explanation lies in Magentic-One's system design, which often involves extended exploratory and re-planning cycles. 
Early-phase errors, such as incorrect task decomposition, improper subtask assignment, or flawed initial assumptions, may not manifest immediately and only become visible after subsequent execution fails. In contrast, mid- to late-phase failures often involve tangible inconsistencies, evidence conflicts, or clear execution failures (e.g., mismatched web information), which provide stronger, more localized signals for attribution.

\subsubsection{{\tool} vs. Who\&When}
\label{sec_result_whowhen}
\textbf{Performance under the Output-only Setting.}
Table \ref{tab:observability-ablation} demonstrates the performance under \textit{All-at-Once w/o metadata\&input}, i.e., only utilizing the step output fields for failure attribution, which is exactly the case as Who\&When benchmark \cite{zhang2025agent}. 
We can see that, solely relying on the output fields, the performance suffers a notable degradation, from 62\% to 51\% in agent-level accuracy and from 28\% to 16\% in step-level accuracy. 
These results are largely consistent with those reported in Who\&When (with the same backbone LLM), i.e., agent-level accuracy being 51.1\% to 54.3\% and step-level accuracy being 12.5\% to 13.5\%.
The key observability gap is therefore not whether intermediate reasoning text appears in the transcript, but whether the actual decision context of each LLM call is recorded.
Who\&When may include intermediate agent outputs, reasoning snippets, or tool-call descriptions, but it does not provide the role-specific prompt, the exact visible history, system-constructed context, agent configuration, or tool/environment information injected into the next prompt.
Without these input-side fields, a later output sequence only shows chronological order, not what information each component actually observed when making its decision.
{\tool} records these input contexts explicitly, enabling attribution methods to distinguish failures caused by missing or transformed upstream information from failures caused by the acting component's own reasoning.

\textbf{Variation in Prompting Strategies.}
A key difference lies in the comparison across prompting strategies: in our experiments, both agent-level and step-level accuracy achieve the best results under the All-at-Once setting, whereas in Who\&When, step-level accuracy peaks under the Step-by-Step setting. 
This discrepancy may stem from the fact that our execution traces contain more steps (average LLM invocations: 20.5 for Captain-Agent, 29.3 for Magentic-One vs. 9.6 and 28.8 in Who\&When), leading to overly long contextual inputs in later interactions of incremental prompts, i.e., exceeding the effective context window of the LLM and thus degrading performance.

We provide more analysis on Appendix \ref{app:whowhen_analysis}.

\subsubsection{Implications and Takeaways}

Our benchmark and experimental findings offer several actionable insights for the future of failure attribution in MAS.

\textbf{Architecture-Aware Attribution.} 
As revealed in Section \ref{sec_failure_patterns}, the types of failure responsible agents and the distribution of decisive failure steps vary significantly across different MAS architectures.
The performance of automated attribution itself is highly dependent on agent types and step positions. 
In practice, this calls for architecture-aware attribution methods that leverage prior knowledge about an MAS's design (e.g., centralized vs. dynamic team formation, tool-heavy vs. planning-heavy workflows) to focus attention on its most vulnerable components and interaction patterns, thereby improving both the efficiency and accuracy of failure attribution.

\textbf{Enhancing Static Attribution with Advanced Reasoning.} Our \textit{Static Agentic} method, while effective, currently employs only basic tool use (e.g., inspecting step I/O, see details in Appendix \ref{app:evaluation_design}). Future work can integrate more sophisticated reasoning strategies, such as explicit hypothesis generation and testing cycles, or graph-based reasoning over extracted agent interaction networks. This could help the agent better synthesize long-horizon dependencies and ambiguous failure patterns present in static traces.

\textbf{Leveraging Dynamic Environment for Deeper Analysis.} The current dynamic configuration primarily performs single-step replay and counterfactual checks, which already improve step-level accuracy. The provided executable environment enables more advanced analysis techniques, such as reading code to infer potentially faulty nodes in the system design, reconstructing the actual control-flow structure from execution traces, conducting systematic state-space exploration around failure points, testing system robustness through automated fault injection, or applying causal discovery algorithms that actively intervene on multiple variables to isolate root causes. These measures allow researchers to better understand how error traces arise. As a result, the benchmark is transformed from a passive dataset into an active laboratory for debugging research.

\textbf{Specializing Models for Attribution via Fine-Tuning.} The strong correlation between model reasoning capability (e.g., Claude-4.5, DeepSeek-R1) and attribution accuracy also highlights the need for specialized models. Promisingly, recent work such as GraphTracer \cite{zhang2025graphtracer} and AgenTracer \cite{zhang2025agentracer} shows that task-specific fine-tuning can enable smaller models to surpass larger base models. 
Future efforts can fine-tune compact models utilizing {\tool}'s traces and running environments, explicitly incorporating structural features (e.g., agent graphs, tool-call sequences) and temporal dependencies as auxiliary training signals to create efficient, attribution-specialized models.

\textbf{Toward Integrated Debugging Tools.} Our findings underscore the practical value of full observability. This motivates the development of integrated debugging tools that automatically capture rich execution traces, visualize agent interaction flows, and suggest potential failure points using attribution models. Embedding such capabilities into MAS development frameworks can significantly reduce debugging overhead.
 
\section{Related Work}
\label{sec:related_work}

\textbf{LLM-based Multi-Agent Systems.}
LLM-based Multi-Agent Systems (MASs) have been widely studied as a paradigm for solving complex tasks, where agents interact through natural language, planning modules, and tool usage to perform tasks such as software engineering, question answering, and decision making \cite{chen2024survey, maldonado2024multi, zhang2025survey}. 
Existing work has explored different system architectures, including centralized planning, decentralized coordination, and hybrid designs, as well as mechanisms for role assignment and communication \cite{wu2024autogen, li2023camel, fourney2024magenticonegeneralistmultiagentsolving, song2025captainagent, 2025mirothinker, jin2025evolution}.
As MAS are deployed in increasingly complex and long-horizon tasks, failures become more difficult to diagnose and failure attribution is even challenging.

\textbf{Failure Attribution in MAS.}
There are several techniques for failure attribution in MASs.
For example, FAMAS \cite{ge2025spectrumanalysis} conducted spectrum analysis on multiple execution trajectories collected by repeatedly replaying a failed task.
AgentTracer \cite{zhang2025agentracer} proposed a lightweight model trained via multi-granular reinforcement learning to jointly optimize step-level and agent-level attribution accuracy.
GraphTracer \cite{zhang2025graphtracer} modeled agent interactions as information dependency graphs, and traced causal information flows. 
ECHO \cite{banerjee2025ECHO} combined hierarchical context representation, objective analysis-based evaluation, and consensus voting to improve failure attribution accuracy.
To reliably evaluate these existing techniques and support the development of new ones, high-quality benchmarks like ours are essential.

\section{Conclusion}
\label{sec:conclusion}

The inherent complexity and non-deterministic interactions in LLM-based MASs make failure attribution a core challenge for ensuring operational reliability and facilitating targeted debugging.
To support this, we introduce {\tool}, a benchmark for failure attribution of LLM-based MASs under full execution observability. 
Experiments show that full observability significantly improves attribution performance, and dynamic replay further enhances attribution capability.
This work lays a practical foundation for future failure attribution research and promotes evaluation practices that mirror real-world debugging.

\section{Limitations}
\label{sec:limitations}
This work focuses on failure attribution under developer-facing settings using execution traces collected from a limited set of representative multi-agent systems. While this scope does not cover all possible system architectures or black-box usage scenarios, it reflects realistic debugging conditions in which full execution traces are available.
As this study is conducted on only three MASs, some of our findings may not generalize to all existing or future systems. However, the selected systems, i.e., Captain-Agent, Magentic-One, and SWE-Agent, are intentionally diverse in their design paradigms, covering dynamic team assembly, centralized orchestration, and specialized software engineering workflows. This deliberate diversity enhances the representativeness of our benchmark and mitigates the risk of architecture-specific bias.

\bibliography{custom}

\appendix

\definecolor{jsonbg}{RGB}{245,246,247}
\definecolor{jsonkey}{RGB}{140,90,40}
\definecolor{jsonstring}{RGB}{70,120,60}
\definecolor{jsoncomment}{RGB}{120,120,120}

\lstdefinelanguage{json}{
  basicstyle=\ttfamily\small,
  backgroundcolor=\color{jsonbg},
  showstringspaces=false,
  breaklines=true,
  frame=none,
  stringstyle=\color{jsonstring},
  keywordstyle=\color{jsonkey},
  commentstyle=\color{jsoncomment},
  morestring=[b]",
}

\newtcolorbox{jsonbox}{
  colback=jsonbg,
  colframe=black!15,
  arc=4pt,
  boxrule=0.5pt,
  left=6pt,
  right=6pt,
  top=6pt,
  bottom=6pt
}

\section{Appendix}
\label{sec:appendix}

\subsection{Detailed Problem Definition}
\label{app:problem_definition}
\paragraph{Multi-Agent Execution and Observability.}
We consider an LLM-based multi-agent system (MAS) $\mathcal{M}$ composed of a finite set of agents
$\mathcal{A} = \{a_1, a_2, \dots, a_N\}$ that collaboratively perform a task $\tau$.
The system executes in discrete steps. At each step, a single agent is selected to act,
possibly through a dispatcher, router, or planning component that is part of the system.
At step $t$, the acting agent is selected as $a(t) = g(h_{t-1})$, where $g(\cdot)$ represents the system-level agent selection mechanism and
$h_{t-1}$ denotes the system context at step $t-1$, summarizing the execution history.
The selected agent receives an input $x_t = \phi(h_{t-1}),$ and produces an output $ y_t = \pi_{a(t)}(x_t),$
where $\pi_{a(t)}$ denotes the policy of agent $a(t)$.
The system context is then updated as $h_t = u(h_{t-1}, a(t), x_t, y_t).$
The execution terminates after $T$ steps, producing an execution trace $\mathcal{T} = \langle s_1, s_2, \dots, s_T \rangle$.
Each execution step $s_t$ is associated with a structured observable record
$
o_t =
\big(
x_t,\;
y_t,\;
a(t),\;
\textit{step\_id}_t,\;
\textit{agent\_id}_t\;
\big)
$.

\paragraph{Task Outcome and Failure Attribution.}
Given a completed execution trace $\mathcal{T}$, the system produces a task outcome.
We define a task outcome function $ \Omega(\mathcal{T}) \in \{0,1\},$
where $\Omega(\mathcal{T}) = 1$ indicates successful task completion and
$\Omega(\mathcal{T}) = 0$ indicates failure.
Failure attribution aims to identify the agent and the execution step responsible for a task failure based on observable execution traces.
We define failure attribution at two levels.
\textbf{Step-level attribution.}
Let $\mathcal{T}_{\le t}$ denote the prefix of the execution trace up to step $t$.
We define the set of feasible continuations from $\mathcal{T}_{\le t}$ under system
$\mathcal{M}$ as
$
\mathcal{C}(\mathcal{T}_{\le t}) =
\{\mathcal{T}' \mid \mathcal{T}' \text{ is a valid continuation from } \mathcal{T}_{\le t}\}.
$
Failure is said to become inevitable at step $t$ if all feasible continuations
result in failure:
$
\forall \mathcal{T}' \in \mathcal{C}(\mathcal{T}_{\le t}),\;
\Omega(\mathcal{T}_{\le t} \oplus \mathcal{T}') = 0.
$
The failure step is defined as the earliest inevitable step
$t^{\ast} = \min \Big\{ t \in \{1,\dots,T\} \mid
\forall \mathcal{T}' \in \mathcal{C}(\mathcal{T}_{\le t}),\;
\Omega(\mathcal{T}_{\le t} \oplus \mathcal{T}') = 0 \Big\}.$
\textbf{Agent-level attribution.}
Given an execution trace $\mathcal{T}$ and its failure step $t^{\ast}$,
agent-level attribution identifies the agent whose action at the failure step
gives rise to the task failure.
Formally, the failure-responsible agent is defined as
$a^{\ast} =
\arg\max_{a \in \mathcal{A}}
\mathbb{I}\!\left[
\exists\, s_{t^{\ast}} \in \mathcal{T}
\;\text{s.t.}\;
a(t^{\ast}) = a
\right],$
where $a(t^{\ast})$ denotes the agent selected by the system to act at step $t^{\ast}$ in the execution trace $\mathcal{T}$.

\subsection{Details about Data Sources} 
\label{app:dataset_sources}

\textbf{LLM-based Multi-Agent Systems.}
We collect execution traces from three representative agentic systems: \textit{Captain-Agent}, \textit{Magentic-One}, and \textit{SWE-Agent}.
Captain-Agent \citep{song2025captainagent} follows an automated system construction paradigm, where a meta-controller dynamically assembles and coordinates a team of agents for each task through iterative planning and execution. In contrast, Magentic-One \citep{fourney2024magenticonegeneralistmultiagentsolving} adopts a manually specified architecture, in which an orchestrator governs a fixed set of specialized agents. 
SWE-Agent \citep{yang2024sweagent} represents a single-agent tool-centric scaffold for software engineering tasks, where planning, orchestration, and tool invocation act as functional components for attribution.
For each system, we use the official implementation and retain the original agent definitions, prompts, and tool interfaces.

\textbf{Task Sources.}
For both \textit{Captain-Agent} and \textit{Magentic-One}, execution traces are collected on tasks from \textit{GAIA} and \textit{AssistantBench}, reflecting general-purpose agentic problem solving involving multi-step reasoning, information gathering, and tool use. \textit{GAIA} tasks typically require agents to decompose a question, retrieve information from multiple sources, and integrate intermediate results through a sequence of reasoning steps. \textit{AssistantBench} tasks further emphasize interactive tool use and sequential decision making, where errors may arise from incorrect tool selection, incomplete information propagation, or early planning mistakes. For \textit{SWE-Agent}, traces are collected on \textit{SWE-Bench Verified} tasks, which focus on software engineering scenarios such as code navigation, modification, and validation. These tasks often involve long execution traces with repeated interactions between language model agents and external development environments. Failures in this setting commonly manifest as incorrect localization of code changes, incomplete fixes, or mismatches between generated patches and test requirements, resulting in qualitatively different failure patterns compared to general-purpose reasoning tasks.

\textbf{Run Configuration.}
For each system-task pair, we execute the system under a fixed configuration $\theta$ and assign a unique run identifier $\rho$. To ensure stable tool invocation behavior while retaining a controlled degree of stochastic diversity, we set the decoding temperature as 0.3 across all runs and repeat each system-task pair for multiple trials.  

\textbf{Trace Statistics.}
Across all systems and task sources, a total of 380 execution traces are collected after data cleaning, among which 220 traces result in task failures and are included as benchmark instances. The number of failed traces varies across systems and task sources, reflecting differences in task difficulty, execution length, domain constraints, and system specialization. Rather than enforcing a uniform failure rate, {\tool} preserves the naturally occurring failure distributions observed during execution. This design choice allows failure attribution methods to be evaluated under realistic conditions, where failures may stem from diverse causes and arise at different stages of execution. 
Each failed run is treated as an independent attribution instance, capturing variability introduced by language model decoding and tool interactions.

\subsection{Details about Trace Collection}
\label{app:trace_adapter}

To enable systematic failure attribution across heterogeneous MAS, {\tool} adopts an automated trace collection pipeline that adapts different systems into a unified execution logging interface without modifying their original implementations.

\textbf{LLM API Middleware.} At the core of the trace collection pipeline is a lightweight LLM API middleware that mediates all interactions between a running MAS and its underlying LLMs. 
Instead of directly communicating with an external LLM service, the MAS is configured to send LLM requests through a middleware. 
It serves as a transparent proxy between a running multi-agent system and its underlying large language model. All LLM requests issued by agents are intercepted by the middleware, forwarded to a user-specified backend model, and logged together with the corresponding responses. This design enables complete observability of LLM-mediated decision making without modifying agent logic, prompts, or system control flow. Instead of directly communicating with an external LLM service, the multi-agent system is configured to route all LLM requests through the middleware. For each request, the middleware records the request payload (including messages and decoding parameters), forwards it to the backend model, receives the response, and returns the response to the calling agent. Since tool invocations in modern MASs are typically triggered by LLM-generated outputs, the middleware additionally captures LLM-mediated tool interactions when such information is observable at runtime. This includes the selected tool name, tool input arguments, and execution results. As a result, both language-model reasoning and subsequent tool-mediated actions are recorded in a unified, system-agnostic manner.

\textbf{Trace Pre-processing.}
We apply lightweight pre-processing to raw traces to expose necessary information in a structured manner while preserving their original form as much as possible. Specifically, we use regular-expression matching to perform two annotations: (1) extracting and tagging agent names in the traces, and (2) categorizing step outputs as either plain LLM responses or tool-mediated interactions. 

\textit{Agent Name Extraction.} Agent names are extracted directly from execution-time observations using pattern matching, without introducing additional semantic inference. In many MAS implementations, agent identities are explicitly referenced in message headers or system-generated markers.
For example, given the message:
\begin{quote}
\texttt{Planner: We need to search for relevant documentation.}
\end{quote}
The agent name \texttt{Planner} is extracted and associated with the current execution step.

\textit{Output type categorization.} Each step output is categorized as either a plain LLM response or a tool-mediated interaction. This distinction is determined by matching tool invocation patterns in the output text.
For instance, the output
\begin{quote}
\texttt{Action: search\_web\\ Input: ``Python regex example''}
\end{quote}
is classified as a tool-mediated step.

\textit{Step and Agent Indexing.} Each recorded trace is assigned a monotonically increasing \textit{step\_id} according to its chronological order in the execution trace.
Agent identifiers (\textit{agent\_id}) are assigned based on the first occurrence of each agent name and remain consistent within a trace.
Each step is linked to both the corresponding \textit{agent\_id} and the \textit{agent name}.

\subsection{Example Execution Trace}
\label{app:trace_example}

We provide a concrete example to illustrate how execution traces are recorded in {\tool}.
The example corresponds to a general travel-planning task executed by a multi-agent system.

\paragraph{Task.}
\textit{``How should a first-time visitor plan a 3-day trip to Walt Disney World in Florida?''}
This task requires multi-step reasoning, external information retrieval,
and verification across multiple constraints,
making it representative of realistic agentic workloads.

\paragraph{Trace Metadata.}
Each execution trace begins with trace-level metadata that provides global context for the execution.
\begin{jsonbox}
\begin{lstlisting}[language=json]
{
  "trace_metadata": {
    "task_id": "WDW-TRAVEL-001",
    "task_instruction": "How should a first-time visitor plan a 3-day trip to Walt Disney World in Florida?",
    "system_name": "Magentic-One",
    "agent_configuration": {
      "agents": [
        "PlannerAgent",
        "SearchAgent",
        "ItineraryAgent",
        "VerifierAgent"
      ],
      "prompts": ["..."],
      "tools": ["web_search"]
    },
    "system_architecture": {
      "description": "A centrally orchestrated multi-agent system with role-specialized agents.",
      "code": ["orchestrator.py", "agents/*.py"]
    }
  }
}
\end{lstlisting}
\end{jsonbox}

The metadata records the task, system identity, agent composition, and architectural information. 
Besides for failure attribution, these fields can also provide essential context for interpreting execution behavior and comparing traces across systems.

\paragraph{Step-level Records.} The execution trace consists of an ordered sequence of step-level records,
each corresponding to a single agent action.
\textit{Step 1: Task Decomposition}
\begin{jsonbox}
\begin{lstlisting}[language=json]
{
  "step_id": 1,
  "agent_id": 0,
  "agent_name": "PlannerAgent",
  "input_context": "Task: Plan a three-day Walt Disney World visit for a family.",
  "output_content": "I will break the task into park selection, ticket options, and visit order."
}
\end{lstlisting}
\end{jsonbox}
The planner interprets the user request and decomposes it into actionable sub-tasks.

\textit{Step 2: Identifying Disney Theme Parks}
\begin{jsonbox}
\begin{lstlisting}[language=json]
{
  "step_id": 2,
  "agent_id": 1,
  "agent_name": "SearchAgent",
  "input_context": "Find the main theme parks at Walt Disney World.",
  "output_content": "Searching official Disney sources.",
  "tool_logs": {
    "tool_name": "web_search",
    "input_arguments": "Walt Disney World theme parks",
    "outputs": "Magic Kingdom, EPCOT, Hollywood Studios, and Animal Kingdom.",
    "execution_status": "success"
  }
}
\end{lstlisting}
\end{jsonbox}
This step involves an external tool invocation.
The raw tool interaction is preserved verbatim in \textit{tool\_logs}. Please note that for each step of the output, there must be an \textit{output\_content}, and only when there is a tool call will there be a \textit{tool\_logs}.

\textit{Step 3: Retrieving Ticket Options}
\begin{jsonbox}
\begin{lstlisting}[language=json]
{
  "step_id": 3,
  "agent_id": 1,
  "agent_name": "SearchAgent",
  "input_context": "Find ticket options for a three-day Disney World visit.",
  "output_content": "Searching ticket types and pricing.",
  "tool_logs": {
    "tool_name": "web_search",
    "input_arguments": "Disney World three-day ticket options",
    "outputs": "Standard tickets allow one park per day; park-hopper tickets allow multiple parks per day.",
    "execution_status": "success"
  }
}
\end{lstlisting}
\end{jsonbox}
Here, the same agent performs another retrieval step,
producing additional evidence required for planning.

\textit{Step 4: Drafting an Initial Visit Plan}
\begin{jsonbox}
\begin{lstlisting}[language=json]
{
  "step_id": 4,
  "agent_id": 0,
  "agent_name": "PlannerAgent",
  "input_context": "Theme parks and ticket options have been retrieved.",
  "output_content": "Draft plan: Day 1 Magic Kingdom, Day 2 EPCOT, Day 3 Hollywood Studios."
}
\end{lstlisting}
\end{jsonbox}
The planner synthesizes retrieved information into a tentative plan.

\textit{Step 5: Feasibility Verification}
\begin{jsonbox}
\begin{lstlisting}[language=json]
{
  "step_id": 5,
  "agent_id": 2,
  "agent_name": "VerifierAgent",
  "input_context": "Verify whether the proposed three-day plan is feasible for a family.",
  "output_content": "The plan is feasible. Park-hopper tickets are optional but not required."
}
\end{lstlisting}
\end{jsonbox}
The verifier checks internal consistency and practical feasibility.

\textit{Step 6: Final Recommendation}
\begin{jsonbox}
\begin{lstlisting}[language=json]
{
  "step_id": 6,
  "agent_id": 0,
  "agent_name": "PlannerAgent",
  "input_context": "Verification completed.",
  "output_content": "Final recommendation: Visit Magic Kingdom, EPCOT, and Hollywood Studios on separate days with standard three-day tickets."
}
\end{lstlisting}
\end{jsonbox}
This final step produces the consolidated answer returned to the user.

This example demonstrates how {\tool} records:
(1) rich trace metadata, (2) fine-grained step-level inputs and outputs, and (3) raw tool interactions,
all in a unified JSON format.


\subsection{Failure Attribution Annotation}
\label{app:annotation}
This section describes the human annotation process used to obtain failure attribution labels, i.e., (1) the agent primarily responsible for the failure, and
(2) the execution step where the failure originates.

\textbf{Annotators.}
The annotation is conducted by three human annotators with at least one year of experience in developing and debugging MASs. The annotators are familiar with agent pipelines, tool-mediated execution, and trace-based analysis.

\textbf{Annotation principle.}
The annotation process is conducted in three rounds.
In the first round, failed execution traces from all multi-agent systems are evenly distributed among the three annotators. Each annotator independently inspects the complete execution trace, including agent actions and tool interactions. 
For each trace, the annotator identifies the failure-responsible agent and decisive failure step. 
Annotators additionally indicate whether they are confident or uncertain about each annotation.

In the second round, all traces marked as uncertain by at least one annotator are jointly reviewed. Annotators collaboratively examine the execution timeline, discuss alternative interpretations of the trace, and reconcile differences through discussion. Consensus is reached through mutual agreement rather than majority voting, and annotations are finalized only when all annotators agree on the attribution outcome.

In the final round, each annotator reviews a subset of annotations produced by the other annotators
to assess consistency with the shared annotation standards. When inconsistencies or ambiguities are identified, the corresponding traces are revisited and re-annotated through further discussion until a consistent labeling standard is achieved across.
This multi-round procedure follows established practices for aggregating expert judgments to approximate reliable ground truth \cite{krippendorff2018content}.
We compute Krippendorff's alpha on the first-round independent annotations, obtaining 0.72 for agent-level labels and 0.64 for step-level labels.
After the consensus process, all finalized labels reflect unanimous agreement among the annotators.

\subsection{Evaluation Design}
\label{app:evaluation_design}

\subsubsection{Evaluation Metrics.}
Given an execution trace $\mathcal{O}(\mathcal{T})$, the failure attribution method is tasked with predicting the failure-responsible agent $\hat{a}$ and the decisive failure step $\hat{t}$. Predictions are evaluated against the ground-truth labels $(a^\ast, t^\ast)$ annotated in our benchmark.
Let $\mathcal{D}$ denote the set of failed execution traces in the benchmark.
We report \textbf{\textit{agent-level accuracy}}, defined as the proportion of traces
for which the predicted agent matches the ground truth:
$\mathrm{Acc}_{\text{agent}} =
\frac{1}{|\mathcal{D}|}
\sum_{(\mathcal{T}, a^\ast, t^\ast) \in \mathcal{D}}
\mathbb{I}[\hat{a} = a^\ast].$
Similarly, we report \textbf{\textit{step-level accuracy}}, defined as:
$\mathrm{Acc}_{\text{step}} =
\frac{1}{|\mathcal{D}|}
\sum_{(\mathcal{T}, a^\ast, t^\ast) \in \mathcal{D}}
\mathbb{I}[\hat{t} = t^\ast].$
Following prior work, we additionally report \textbf{\textit{step-level accuracy with tolerance}},
where a prediction is considered correct if $|\hat{t} - t^\ast| \leq \delta$,
with $\delta$ fixed across all experiments.
All results are averaged over 3 independent runs.

\subsubsection{Observability Configurations.} 
We evaluate two categories of observability configurations.
For \textbf{\textit{Static}} configuration, we provide the overall trace information as demonstrated in Section \ref{sec:trace_fields} for failure attribution. 
For \textbf{\textit{Dynamic}} configuration, besides the static information, we also provide a replayable execution environment for each trace. 
This setting reflects an interactive debugging workflow where static inspection is complemented by execution-based validation.

We additionally utilize \textbf{\textit{Static w/o metadata}}, \textbf{\textit{Static w/o input}}, and \textbf{\textit{Static w/o metadata\&input}} to further evaluate the necessity of different types of information for failure attribution.

\subsubsection{Failure Attribution Techniques.}
\label{app:tech_details}
For above mentioned static configuration, we evaluate three commonly-used LLM-based failure attribution techniques \cite{zhang2025agent,zhang2025graphtracer,ge2025spectrumanalysis} and two popular agent-based techniques \cite{yang2024sweagent}. 
The first three techniques present the failure trace to the LLM, and let the LLM identify the responsible agent and decisive step. 
They differ in how the trace is provided.
\textbf{\textit{All-at-once}} provides the full trace in a single context window and asks the LLM to output the failure-responsible agent and decisive failure step.
\textbf{\textit{Step-by-step}} reveals the trace incrementally and asks the LLM to decide at each step whether it contains the failure origin, terminating when a step is selected.
\textbf{\textit{Binary search}} repeatedly asks the LLM to localize the failure to a sub-range of steps, halving the search space until a single step is identified.

For the first agent-based technique (short for \textbf{\textit{Static Agentic}}), we adopt mini-SWE-agent, which can navigate the trace information, retrieve related fields as needed, and make a conclusion gradually.
Specifically, we modify the codebase of \emph{mini-SWE-agent} by adding three tools:
(1) inspecting the outputs of all executed steps;
(2) inspecting the input and output of a specific step; and
(3) submitting the final answer.
These extensions enable the LLM agent to flexibly and dynamically switch between global information and fine-grained local details during reasoning.

For \textbf{\textit{Dynamic Agentic}}, the method first proposes candidate failure attributions derived from the Static Agentic technique, including candidate steps and agents, and then re-runs the system from the corresponding execution point to issue counterfactual checks.
Specifically, we first leverage the Static Agentic method to infer up to $n$ ($n \leq 3$) candidate \emph{mistake agent/step/reason} triples, together with their corresponding \emph{expected oracle}.
The expected oracle represents the anticipated output of the identified step if the specified mistake reason had not occurred.
It is inferred by the Static Agentic model as part of the candidate attribution hypothesis, rather than provided by human annotators or derived from the ground-truth labels.
We then replay the task trajectory, and when execution reaches the identified mistake step, we intervene by modifying the input request of that step through an LLM API middleware.
This intervention guides the agent to avoid the error induced by the mistake reason.
We subsequently observe whether the next $k=3$ steps satisfy the expected oracle and whether the previously observed failure modes reoccur or the agent deviates from the task objective.

We do not execute the entire task to completion; instead, we restrict our observation to $k=3$ steps.
This design choice is motivated by our empirical finding that many task failures stem from systemic design issues, where even corrective interventions may not guarantee success over long horizons.
By focusing on a limited window of $k=3$ steps, we assess local behavioral improvements within a bounded range, which provides a more reliable basis for diagnosis and judgment.
Thus, Dynamic Agentic validates local responsibility for a candidate step, rather than attempting to repair the entire system or prove global causal sufficiency.
We choose $k=3$ because shorter windows often fail to expose meaningful behavioral changes, while longer windows introduce increasing variance from downstream stochasticity, tool noise, and unrelated planning effects.
The decoding temperature is fixed at 0.3 to preserve local trajectory stability while still allowing limited behavioral adaptation after the intervention.

Note that there are other techniques for failure attribution in MASs (i.e., ECHO \cite{banerjee2025ECHO}, AgenTracer \citep{zhang2025agentracer}, GraphTracer \citep{zhang2025graphtracer}, and FAMAS \cite{ge2025spectrumanalysis}). 
However, they don't release runnable implementations or source code. 
We attempt to reproduce them, but couldn't obtain performance consistent with the numbers reported in original works. 
Therefore, we don't include them in evaluation.

\subsubsection{Application Scenarios.}
We evaluate attribution under two settings that simulate realistic debugging demands.
In the \textbf{\textit{With ground truth}} scenario, attribution methods are provided with task-level reference information, which is commonly available in development or diagnostic scenarios.
Specifically, for GAIA and AssistantBench, we use the official task ground-truth answers as references.
For SWE-Bench-Verified, where explicit reference answers are not available, we instead treat the observed test outcomes of the evaluation instances (i.e., pass/fail signals from the test suite) as ground-truth supervision.
In contrast, in the \textbf{\textit{Without ground truth}} scenario, the attribution method relies solely on the execution trace (and optional replay information), reflecting practical debugging scenarios where no definitive reference outcome is accessible.

\subsubsection{LLMs Under Evaluation.}
We evaluate failure attribution performance across a set of representative LLMs, covering both closed-source and open-source models. 
We use GPT-4o \citep{hurst2024gpt} as the default LLM in the main experiments. We additionally compare against GPT-4o-mini \citep{openai_gpt4o_mini_2024}, Claude-4.5-Sonnet \citep{anthropic_sonnet4_5_2025} (closed-source) and Qwen3 \citep{yang2025qwen3}, GPT-OSS-20B \citep{agarwal2025gpt}, DeepSeek-R1 \citep{guo2025deepseek} (open-source), to assess whether the findings are consistent across model families and capability levels. 

\subsection{Analysis on Who\&When Benchmark}
\label{app:whowhen_analysis}

\begin{figure}[!t]
\centering
\vspace{0.1in}
\includegraphics[width=0.47\textwidth]{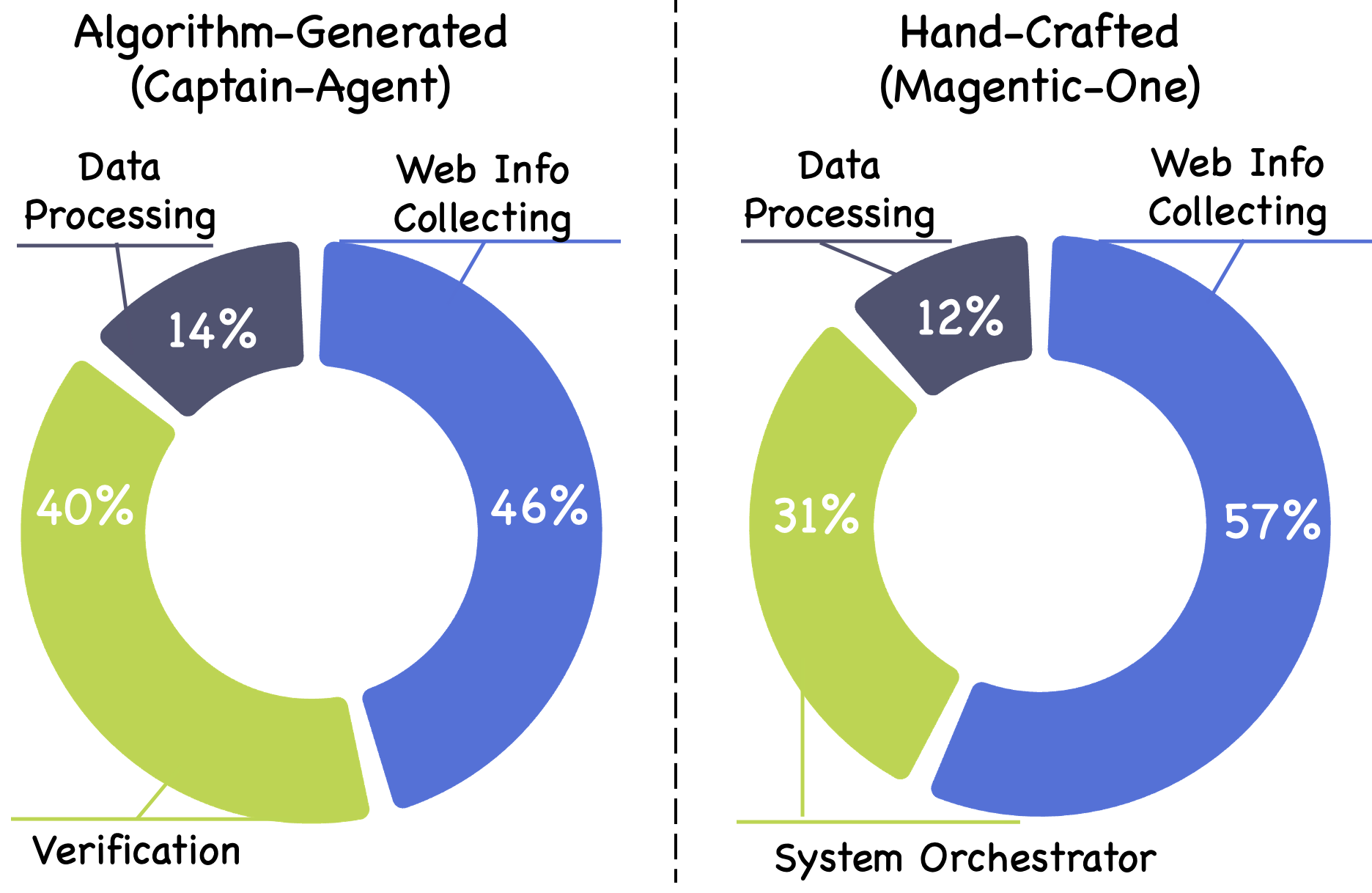}
\caption{Distribution of failure agent in Who\&When.}
\label{fig:wwdistribution}
\end{figure}

\begin{figure}[!t]
\centering
\includegraphics[width=0.47\textwidth]{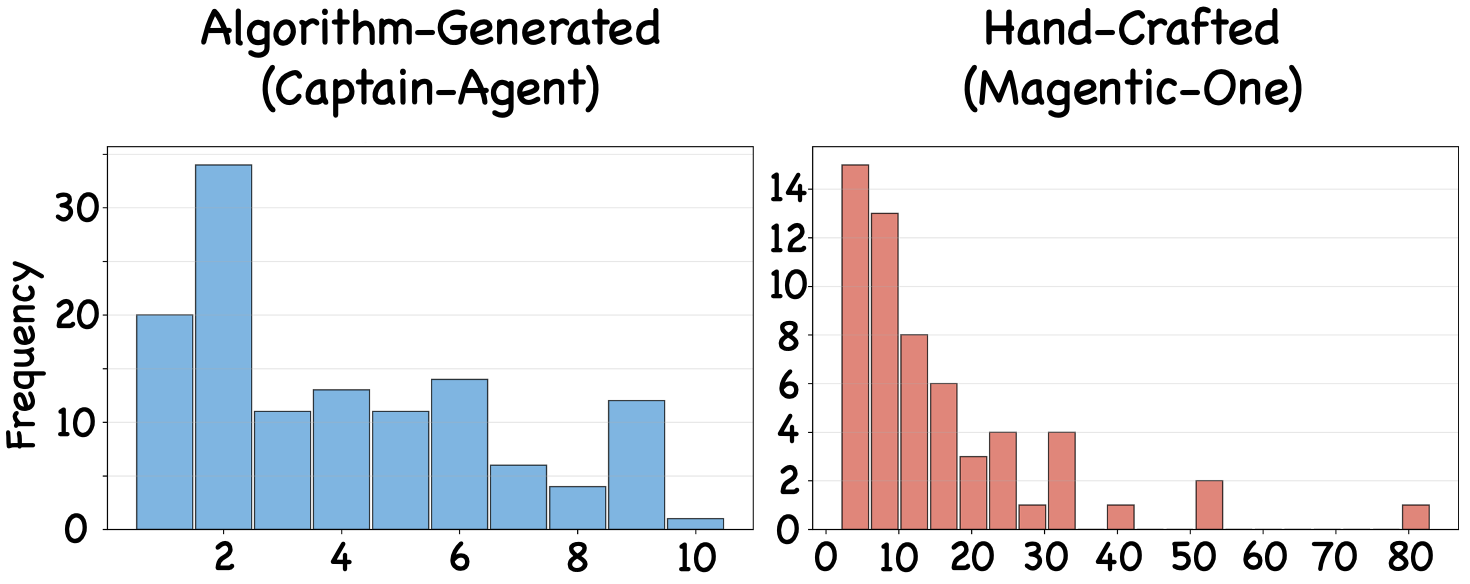}
\caption{Distribution of failure step in Who\&When.}
\label{fig:wwdifferent}
\end{figure}

We report the distribution of failure agent and failure step in Who\&When Benchmark in Figure \ref{fig:wwdistribution} and \ref{fig:wwdifferent}. 
These trends align closely with those observed in {\tool} (see Figure \ref{fig:failure_agent} and \ref{fig:failure_step}).

The slight difference lies in the average number of LLM invocations: in Who\&When, Captain-Agent and Magentic-One respectively have an average of 9.6 and 28.8 calls, while these figures for {\tool} are 20.5 and 29.3. 

It is also worth noting that the step count for Magentic-One appears significantly higher (almost 80 steps) in Who\&When as shown in Figure \ref{fig:wwdifferent}; this discrepancy stems from differences in step-counting strategies between the two benchmarks.
For example, CaptainAgent omits the scheduling process among speaking agents, and ComputerTerminal executes code from the previous step directly via the command line without invoking an LLM. In Magentic-One, the process of reasoning about and selecting the next speaking agent is decomposed into multiple steps.
We reconstructed the actual number of LLM invocations in the Who\&When dataset: Algorithm-Generated (Captain-Agent) averages 9.6 invocations, Hand-Crafted (Magentic-One) averages 28.8, and the overall average is 15.7.

\subsection{Privacy Considerations}
\label{sec:privacy}
{\tool} is constructed using execution traces that include step-level inputs and outputs of agents. While such information is essential for fine-grained failure attribution, it may contain sensitive content, depending on the task domain and system configuration. For example, execution traces may include user-provided instructions, intermediate reasoning artifacts, or tool invocation details that could expose proprietary logic or private information.

To reduce potential privacy risks, we adopt several conservative design choices during benchmark construction. First, all traces are collected from controlled task executions over public benchmark sources, rather than from real user interactions. Tasks are predefined and do not involve personal user data. Second, for web-based executions, we use an isolated Playwright-controlled Chromium instance without personal browsing profiles, cookies, or user-specific history. Third, we apply automatic filtering to remove local file paths and potential personal identifiers from collected traces, followed by manual inspection of each released trace. Fourth, when releasing the benchmark, sensitive fields can be selectively anonymized or transformed, for example by replacing raw inputs or outputs with hashed or redacted variants when appropriate.

\subsection{Human Annotation Protocol and Ethics Statement}
\label{sec:ethics_annotation}

We employed three expert annotators with prior experience in developing and debugging LLM-based multi-agent systems to provide failure attribution labels for the benchmark. Each annotator was instructed to identify (i) the failure-responsible agent and (ii) the earliest execution step at which task failure becomes inevitable, based on the complete execution trace, including agent inputs, outputs, and tool interactions.

The annotation guidelines emphasized consistency, trace-based reasoning, and adherence to the formal definition of step-level and agent-level failure attribution described in Appendix~\ref{sec:problem_definition}. Annotators conducted their analysis independently in the first round, followed by joint discussion and reconciliation for cases marked as uncertain, until consensus was reached.

All annotators participated voluntarily and were fully informed of the purpose of the study and the intended use of the annotated data for research and benchmarking purposes. The annotation task involved only synthetic tasks or publicly available benchmark data (GAIA, AssistantBench, and SWE-Bench) and did not include any personal, sensitive, or user-generated data.

No external recruitment or crowdsourcing platforms were used, and no demographic data were collected. As the study involved only technical annotation of non-sensitive data by informed research participants, it did not require approval from an institutional ethics review board.

\subsection{Artifact License and Usage Terms}
\label{sec:artifact_license}
{\tool} including execution traces, annotations, and associated metadata, is released for research purposes only. All benchmark artifacts will be made publicly available upon publication through an open-access repository.

The execution traces and annotations are distributed under the Creative Commons Attribution 4.0 International (CC BY 4.0) license, which permits use, distribution, and adaptation for research purposes with appropriate attribution. Before release, the artifacts are filtered and manually reviewed to avoid including personal, sensitive, or proprietary information.

For third-party components referenced in the benchmark, including task sources such as GAIA, AssistantBench, and SWE-Bench, we follow their original licenses and terms of use. Users of TraceElephant are responsible for complying with the licenses of these upstream resources when using or redistributing the benchmark.

The benchmark code, including trace collection scripts and evaluation utilities, will be released under a permissive open-source license, and detailed licensing information will be provided in the accompanying repository.

\end{document}